# Supporting Management of Gestational Diabetes with Comprehensive Self-Tracking: Mixed-Method Study of Wearable Sensors

Mikko Kytö[1,2], Saila Koivusalo[4,5], Heli Tuomonen[2], Lisbeth Strömberg[2], Antti Ruonala[2], Pekka Marttinen[3], Seppo Heinonen[4], Giulio Jacucci[2]

[1] IT Management, Helsinki University Hospital

[2] Department of Computer Science, University of Helsinki

[3] Department of Computer Science, Aalto University, Finland

[4] Department of Obstetrics and Gynecology, Helsinki University Hospital, University of Helsinki

[5] Shared Group Services, Helsinki University Hospital, University of Helsinki

Corresponding author:

Mikko Kytö

mikko.kyto@hus.fi

**Background:** Gestational diabetes (GDM) is an increasing health risk for pregnant women as well as their child. Telehealth interventions targeted to management of GDM have been shown to be effective, but they still have required work from health care professionals for providing guidance and feedback. Feedback from wearable sensors has been suggested to support self-management of GDM, but it is unknown how the self-tracking should be designed in clinical care.
**Objective:** To investigate how to support self-management of GDM with self-tracking of continuous blood glucose and lifestyle factors without the help by health care personnel. We examined comprehensive self-tracking from self-discovery (i.e., learning associations between glucose levels and lifestyle) and user experience perspectives.
**Methods:** We conducted a mixed-methods study where women with GDM (N=10) used a continuous glucose monitor (Medtronic Guardian) and three physical activity sensors: activity bracelet (Garmin Vivosmart 3), hip-worn sensor (UKK Exsed), and electrocardiography sensor (Firstbeat 2) for a week. We collected the data from the sensors, and after the usage participants took part in semi-structured interviews about the wearable sensors. Acceptability of wearable sensors was evaluated with Unified theory acceptance and use of technology (UTAUT) questionnaire. In addition, maternal nutrition data was collected by 3-day food diary, and self-reported physical activity was collected with a logbook.

**Results:** We found that a continuous glucose monitor was the most useful sensor for the self-discovery process, especially when learning associations between glucose and nutrition intake. We identified new challenges of using data from continuous glucose monitor and physical activity sensors in supporting self-discovery in GDM. These challenges included (1) dispersion of glucose and physical activity data in separate applications, (2) missing important trackable features, like amount of light physical activity and other types of physical activity than walking, (3) discrepancy in the data between different wearable physical activity sensors and between continuous glucose monitor and capillary glucose meters, and (4) discrepancy in perceived and measured quantification of physical activity. We found that body placement of sensors as a key factor in the quality of measurements, preference, and ultimately a challenge for collecting data. For example, a wrist-worn sensor was worn more than a hip-worn sensor. In general, the study showed overall a high acceptance for wearable sensors.

**Conclusions:** A mobile application where glucose, nutrition, and physical activity data are combined in a single view is needed to support self-discovery. The design should support tracking features that are important for women with GDM (such as light physical activity) and data for each feature should originate from a single sensor to avoid discrepancy and redundancy. Future work involves evaluating the effect of such a mobile application on clinical outcomes with a larger sample size.

**Clinical Trials:** clinicaltrials.gov (clinical trial reg. no. NCT03941652)

# 1    Introduction

Gestational diabetes (GDM), defined as hyperglycemia first recognized during pregnancy, is an increasing global challenge affecting currently approx. 8%-23% of pregnancies depending on the continent [1]. GDM has considerable health effects as it increases both mother's and child's risk for short- and long-term health disadvantages [2] . Although GDM is a temporary condition that lasts until the birth of the child, GDM increases the later risk of type-2-diabetes for mothers over seven times [3]. Healthy lifestyle choices help in GDM management, nutrition being the primary factor affecting glucose levels [4], but also physical activity [5–9], stress [10], and sleep [11] have an impact on glucose homeostasis. However, women with recently diagnosed GDM do not adequately know how their own lifestyle choices influence glucose levels [12,13], although they need to adapt to the new situation quickly [14]. Given that pregnancy usually lasts approx. 40 weeks and GDM is diagnosed after 12 to 28 weeks of pregnancy, any health intervention designed for managing GDM is used for a limited time (for approximately 12-28 weeks). On the other hand, women with GDM are extra motivated for managing diabetes due to the child [13,15]  and pregnancy represents an exceptional opportunity for lifestyle changes [16].

A recent meta-analysis of eHealth interventions targeted to women with GDM shows that interventions providing weekly or more frequent feedback from health care professionals to women with GDM have shown the potential to improve glycemic control [17]. Typically, in these interventions, women with GDM can communicate with the study interventionists remotely [18,19]. For example, a recent study by Miremberg et al. [18] revealed a statistically significant improvement in glycemic control among women with GDM when systematic feedback was provided by study personnel; every evening the participants received individualized feedback via e-mail from the clinical team regarding their daily glycemic control. However, mHealth interventions without such substantial input from health care professionals are limited and have not shown to be effective [20,21]. We expect that the effectiveness of mHealth interventions can be increased with comprehensive self-tracking through wearable sensors by providing more insight for women with GDM into learning associations between lifestyle and glucose levels [22,23], a process known as self-discovery (e.g.[24]). To establish knowledge on how self-tracking with wearable sensors (including glucose levels and lifestyle) should be designed to support self-management in GDM, we explored the usage of continuous glucose monitor (CGM) and three types of wearable sensors for measuring physical activity. The overall aim was to examine how wearable sensors can support self-discovery and behavior change, and how women with GDM experience them.

## 1.1 Wearable Sensors for Supporting Self-discovery for Women with GDM

Wearable sensors (e.g., fitness trackers) have been included in investigations on the management of non-communicable diseases, such as diabetes, migraine, and multiple sclerosis [25–32]. Also, in pregnancy a recent review shows that wearable sensors have potential to support physical activity among pregnant women, decrease gestational weight gain, predict neonatal outcomes, and support monitoring of fetal heart rate and movements [33]. However, no studies exist where the focus would have been on investigating how different wearable sensors (e.g., in terms of body placement) and their data can support self-discovery. Traditionally, the studies on personal discovery in diabetes management have been based on the data that users enter to an app [34] or write on a paper-based journal [24].

The personal discovery of understanding medical conditions with self-tracking data has gained a lot of attention [24,25,27,29,35–37]. Personal discovery is an iterative and complex process consisting of multiple stages [24,26,35]. These stages include finding potential features that may affect the desired outcome, forming hypotheses, and evaluating their impact on outcome [24,38]. In diabetes, successful self-management requires knowledge of how one's activities and lifestyle (e.g., nutrition, physical activity, sleep, and stress) affect glucose levels. To help people with diabetes in self-discovery, self-tracking with wearable sensors together with glucose monitoring may provide a useful tool. However, it is largely unknown what is the role of self-tracking of activities and lifestyle together with glucose levels with wearable sensors in the self-discovery process. For example, while physical activity and sleep have been found to influence glucose levels [6,8,11] and a handful of wearable sensors for measuring physical activity and sleep are available for self-tracking, the applicability of wearable sensors in supporting of people with diabetes understanding of how their own lifestyle choices affect glucose levels is largely unknown.

Women with GDM represent an interesting user group to study self-discovery, as they have not used to manage their condition for long. The design of supporting the discovery phase becomes an especially important part of the management of GDM, as "coming to terms with GDM" and learning new strategies for self-regulation are important phases in GDM self-management [13,15]. Qualitative studies report feelings of failure, anxiety, loss of control, and powerlessness after receiving a GDM diagnosis [13,14]. However, women with GDM experience "a steep learning curve"; they go from the initial shock of the diagnosis to acceptance and active management of their condition [39].

For women with GDM, it is typical to find associations between nutrition and blood glucose by trying out different foods and measuring glucose afterward [13,39,40]. The behavior where patients try to establish hypotheses between daily activities and changes in disease-specific outcomes has been identified as a stage-based discovery process [24,35,38].

The framework from Mamykina et al. [24] is formulated to explain the discovery process between daily activities on changes in blood glucose levels. According to the framework [24], the self-discovery consists of four stages *1) Feature selection:* individual identifies activities that they believe have an impact on outcome (blood glucose in the context of diabetes), *2) Hypothesis formulation:* individual formulates suspected association with activities and outcome, *3) Hypothesis evaluation:* individuals observe new information about their condition and evaluate how it fits to already collected data, and *4) Goal specification:* individuals formulate future goals based on identified relationships between features and outcome.

Multiple studies emphasize the importance of automatic data collection in diabetes apps [22,41], although this is rarely found in apps used in diabetes research [22,41]. Current standards emphasize the necessity of self-tracking of glucose levels in diabetes management [5], and measurement of blood glucose levels has been found to be the most important feature of a GDM app [42]. However, the requirement of manually entering the blood glucose values has decreased significantly collecting the glucose data [42,43]. Glucose measurements can be done automatically and more frequently with CGMs. CGMs are found to be acceptable among women with GDM [44–47]. However, recent research suggests that CGM alone does not improve glycemic control [45,48] or decrease macrosomia [47]. One reason is that the cause-and-effects between lifestyle choices and glucose levels are not clear for women with GDM after receiving diagnosis [13–15,39,40].

While self-discovery frameworks have been critiqued of expecting too rational and coherent behavior from people using self-tracking [25] (users are not scientists, cf. [49]), the trial-and-error aspect (hypothesis formulation and evaluation) has been identified as typical behavior among women with GDM [13,39,40]. Moreover, the framework by Mamykina [24] also considers the iterative nature of self-discovery, which is important in the context of GDM as the development of pregnancy has an impact on glucose control [50]. Objectively and automatically measured, and constantly available data through wearable sensors data can be expected to support self-discovery [26,27].

## 1.2 User experience with Wearable Sensors for Women with GDM

Self-tracking is often mentioned as an effective behavior change technique [51], for example, shown as an increased physical activity among people with type-2-diabetes [52]. Thus, we investigate the possibilities and challenges of self-tracking with wearable sensors beyond self-discovery. Wearable sensors have the potential to facilitate managing GDM, as there is proof that lifestyle interventions using wearable sensors can be effective among pregnant women. For example, Chan and Chen [53] reported in their review that the interventions with wearable devices for increasing physical activity among pregnant women were more effective than without.

Physical activity is one of the cornerstones in managing GDM [5,7], but the automatic collection of physical activity data has gained minimal attention in GDM apps [22]. This was emphasized in a study by Skar et al. [42], who asked women with GDM to manually enter their physical activity data into an application, but no one did, preventing the collection of any physical activity data. This is understandable, as pregnant women often have limited energy for monitoring their own behavior, since they already have a lot to do and to deal with [40,54]. Rigla et al. [55] enabled tracking of physical activity for women with GDM by recording it with an accelerometer in a mobile phone. However, recording required manual start and stop by pressing buttons in a mobile app, and participants recorded their physical activity only approx. once a week on average. Even engagement with automatic self-tracking has been show to decrease among people with type-2-diabetes and type-1-diabetes [56]. For example, Böhm et al. [56] reported that the number of active users of CGM had dropped over 20% after 20 weeks, and similarly, active users of automatic physical activity tracking had dropped over 30% after 20 weeks.

The other issue to consider in addition to the automaticity of tracking is what types of physical activities should be possible to track. Carolan et al. [15] noted that although walking is commonly advised for women with GDM by diabetes educators and midwives, it can be painful for many. However, automatic self-tracking beyond steps is more challenging. Årsand et al. [30] found that the largest problem for people with type-2-diabetes to track their physical activity was that wearable sensors did not support the measurement of other activities, such as cycling and swimming, which is a common physical activity among pregnant women [57]. More recent studies imply that wearable sensors have still rather low validity in tracking physical activities beyond walking and running, such as bicycling and resistance training [58].

Studies investigating the practical challenges of wearable sensors for self-tracking among women with GDM are largely lacking. As described above there has been only a few studies that have enabled self-tracking of physical activity among women with GDM [42,55], and in the case of self-tracking other lifestyle factors (e.g., sleep and stress) with wearable sensors, no studies exist investigating self-discovery among women with GDM. In the context of pregnancy, automatic self-tracking of lifestyle (e.g. nutrition, physical activity, and symptoms) has been argued to help in countering pregnancy-related health risks [59,60]. However, some women perceive pregnancy medicalization and that they lack control over their own bodies even without multiple wearable sensors [13,54]. The use of sensors can further increase the feeling of losing a normal pregnancy [13]. Moreover, it is unclear how the sensors fit pregnant women, whose physical and mental condition is different from the general population. Pregnancy causes several lifestyle changes (e.g., diet limitations), physical changes (e.g., difficulty to move, contractions of the uterus, increased waist size and heart rate), sleeping disorders, and tiredness. The effect of differences in these conditions on self-tracking with wearable sensors should be investigated.

## 2 Methods

### 2.1 Research design

We conducted a mixed-methods study where women with GDM (N =10) used a variety of wearable sensors and their mobile apps for a week. Our primary aim was to examine how wearable sensors can support self-management of GDM. We studied this with 2 research questions (RQs) as shown in Textbox 1. We investigated how self-tracking with wearable sensors can support/inhibit the self-discovery (RQ1) and how women with GDM experience wearable sensors (RQ2).

Textbox 1. The research questions.

> - RQ1: How self-tracking with wearable sensors (not only CGM) can support/inhibit the self-discovery of women with GDM? We investigate the role of wearable sensors at each stage of the self-discovery process (feature selection, hypothesis formulation, hypothesis evaluation and goal setting) as described in Section 1.1.
> - RQ2: How do women with GDM experience wearable sensors? Although wearable sensors have been investigated with pregnant women and people with diabetes type-1 or type-2, the knowledge how women with GDM perceive wearable sensors is less known as described in Section 1.2.

The study was performed in Finland and in compliance with the Declaration of Helsinki, approved by the Ethics Committees of Helsinki Central Hospital, and registered at clinicaltrials.gov (clinical trial reg. no. NCT03941652).

## 2.2 Sensors

### 2.2.1 Continuous glucose

Medtronic Guardian Connect CGM with Enlite sensor (Medtronic, Ireland; see Figure 1) continuously measures tissue glucose. Flexible filament is inserted just under the skin to measure glucose levels in interstitial fluid every 5 minutes. Values are sent to Medtronic Guardian app via Bluetooth. If Bluetooth connection is not possible, CGM system transmitter collects the data for up to several days. Medtronic requires calibration of the sensor by fingertips blood glucose measurements 2 times a day. The overall mean absolute relative difference has been reported to be 13.6% [61]

The Medtronic CGM was attached to the skin by a study nurse. This was because participants wished to wear the CGM in the arm and they could not attach the CGM to the skin using only one hand.  Currently, CGMs do not allow tracking of lifestyle data and additional sensors are needed to support tracking beyond glucose.

### 2.2.2 Physical activity

We chose to use multiple physical activity sensors to study which sensor or combination of sensors should be used in terms of wearing comfortability and provided data. Please see Figure 1 and Table 1 for details. Exsed (UKK-Institute, Finland) was worn on a hip and provided data about standing,sitting, and standing. The data analysis is based on validated MAD-APE algorithms [62,63]. These analyses have been employed in population-based studies of Finnish adults [64,65].  Vivosmart 3 (Garmin, Switzerland) was worn on a wrist and provided data about intensity minutes. Vivosmart 3 has been shown to measure steps well at slow walking speeds (mean absolute percentage error was 1.0%) [66], which is important as walking speed is affected by pregnancy [67].

The physical activity sensors also varied in terms of how visible they were for nearby others. The Exsed could be worn in a discreet manner so that others would not see it, whereas the Vivosmart 3 on the wrist is more conspicuous. This *physicality* has been shown to be a prominent issue in wearable sensors [49].

Heart rate variability (HRV) sensor Firstbeat Bodyguard 2 (Firstbeat Technologies, Finland) was added to explore the validity of physical activity and sleep data recorded with physical activity sensors. The device is able to continuously measure beat-to-beat heart rate variability with <3ms error and >99.9% detection rate as compared to clinical grade electrocardiography (ECG) [68].

Due to incompatibility issues between different operating systems and sensors, the participants were given iPod touch with the sensor applications pre-installed. The participants were able to use their own mobile phone with Vivosmart 3, as we found no incompatibility issues in the Garmin Connect app prior to the study.

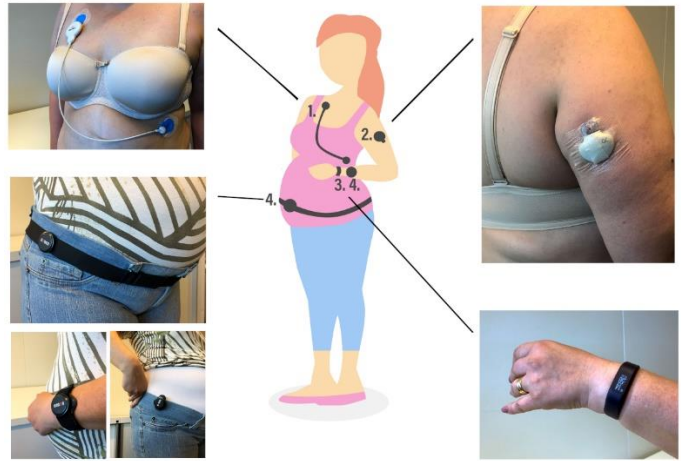

Figure 1. The wearable sensors used in the study: 1) Firstbeat, 2) Medtronic, 3) Vivosmart 3, and 4) Exsed.

## 2.3   Recruitment and data collection

Our goal was to recruit 10 women with GDM to the study from maternity and antenatal clinics in the Helsinki Metropolitan Area (Finland). The goal for the number of participants is similar to multiple qualitative studies on women with GDM [12]. The clinic nurse asked women with GDM at least at 24 gestational weeks about their interest in participation. If interested, the study nurse contacted the mother with more information about the study and confirmed eligibility. Exclusion criteria were type-1 or type-2 diabetes, use of medication that influences glucose metabolism (e.g., oral corticosteroids, metformin, insulin), a GDM diagnosis in previous pregnancies, current substance abuse, severe psychiatric disorder, significant difficulty in cooperating (e.g., inadequate Finnish language skills), and significant physical disabilities that prevent the use of a smartphone or moving without aid. Data were collected using the following procedure. After obtaining informed consent, we collected background information (e.g., age, pregnancy weeks, and familiarity with mobile apps) through a questionnaire.

Participants were asked to wear wearable sensors (see Section 2.2) for six days and nights, after which they were interviewed using their native language (Finnish). The length of the usage period was decided based on the battery life of the transmitter of the CGM, which is six days. To compare data from wearable sensors with their perception of physical activity and sleep, participants filled out a logbook for physical activity and sleeping (duration in hours) for six days, For the physical activity we asked participants to write down the type of activity, duration, and intensity (light, moderate or vigorous). The perceived intensity levels were defined according to descriptions by Norton et al. [69]. Also, Firstbeat used the same intensity categorization as provided in [69]: 20-40% of maximal oxygen consumption ($VO_2$ max) is light physical activity, 40-60% of $VO_2$ max is moderate physical activity, and over 60% of $VO_2$ max is vigorous physical activity. Vivosmart 3 shows intensity of physical activity as intensity minutes, which are gathered when at least 10 consecutive minutes of physical activity at a moderate level is performed. Physical activity at vigorous level doubles the gathered intensity minutes. Explicit thresholds for moderate and vigorous activity were not provided in documentation. Exsed did not provide data regarding intensity of physical activity.

One of the most prominent features is tracking and managing diet, as this is the primary factor that affects glucose levels. However, wearable eating detection systems are not able to detect the macros of the food [70,71]. As such, wearable sensors are not used here to measure diet and participants kept a logbook for diet for three days during the study period. We chose to gather diet data from three days, because keeping a food diary is laborious, and three days have shown to provide valid results [72].

Before starting the measurement period, the participants were met by an experimenter and a study nurse. In the meeting, participants gave written consent, filled in a background questionnaire, and were instructed on how to use the sensors. They were given contact in case they faced problems in using the sensors. Finally, at the end of the meeting, participants filled in a technology acceptance questionnaire based on the Unified Theory of Acceptance and Use of Technology (UTAUT) [73], which has been widely used in evaluating the acceptance of technology in diabetes management [74]. After the usage period, the participant filled the same UTAUT-questionnaire and took part in a semi-structured interview, which was audio-recorded. At first, we asked questions concerning all the sensors, such as how they impacted the users' daily life. After this, we asked questions concerning each sensor, such as what they were able to discover from the data, how the data impacted their daily behavior, what data they valued, and what challenges the users had with each sensor. See Textbox 2 for the main interview questions. Interviews were conducted in quiet places that were easiest for the participants to arrive at and they were conducted in their mother language. Interviews lasted approx. 1 hour on average. After a 15-min break, participants continued in an interview about a prototype GDM application (results are reported elsewhere [23]).

## 2.4 Analysis

Interviews were transcribed and 2 researchers familiarized themselves with the interviews by reading the transcripts. The analysis was done according to the framework method, which is a recommended approach for multidisciplinary health research [75]. We used self-tracking of blood glucose, diet, physical activity, sleep, and stress as initial codes. Coding was implemented with Atlas.ti by employing emergent theme analysis of the data collected [76], resulting in 66 codes altogether. These codes were combined into larger categories, which are subheadings in Section 3.2, Section 3.3, and Section 3.4, and presented and discussed in relation to main themes of the study (i.e., self-discovery and experiences with wearable sensors).

Quotes provided in the results were translated into English intelligent verbatim, a process whereby filler words such as "er" are removed during translations. Log files from the sensors were used to determine how much the participants wore them, how data from the sensors correlated with self-reported data, and whether there were differences in data between the sensors. The statistical significance of differences in data between sensors were computed with Friedman test and correlations between automatically measured and self-reported data were calculated as Spearman or Pearson correlation, depending on the test for normality (Shapiro-Wilk). Finally, we triangulated among these data sources (interviews, data from the sensors, and logbooks) to understand how self-tracking with wearable sensors should be designed to support self-discovery.

**Table 1.** Wearable sensors worn by the participants. Participants wore all the sensors simultaneously.

| Sensor name | Type | Data provided | Wearability (see Figure 1) | Components | User interface (UI) | Waterproof? | Worn by each participant |
|---|---|---|---|---|---|---|---|
| Medtronic Guardian Connect CGM with Enlite sensor | Continuous glucose monitor (CGM) | Interstitial fluid glucose value in every 5 mins. | Typically worn on the area of abdomen which is at least five centimeters from the navel, but participants wished for attaching to the upper arm. | Enlite sensor: flexible filament measures glucose levels in interstitial fluid. Guardian Connect transmitter: Bluetooth | None. Data access through a mobile app (Medtronic Guardian Connect). The app enables viewing time series of glucose values, the viewing range can be changed from one hour to one day. Users insert the calibration values twice a day, and it is possible to add carbohydrates and physical activities to the timeline. | Yes, up to 2.5 meters for up to 30 min. | M = 94% of the time (23h 3 min / day) |
| Garmin Vivosmart 3 | Activity tracker | Steps, intensity minutes, stairs climbed, heart rate, sleep duration, sleep quality, stress, calorie consumption | Worn in a wrist with an adjustable plastic strap | Bluetooth Smart, ANT+, 3D accelerometer, optical heart rate sensor (green LED), barometric altimeter, ambient light sensor | Touch screen, data access through a mobile app (Garmin Connect). The app enables viewing many kinds of information about the recorded data, the time span of the graphs can be varied between one day and one year. | Yes, up to 50 min. | M = 93% of the time (22 h 30 min / day) |
| Exsed | Activity tracker | Duration of physical activity, sedentary behavior, and sleep sensor Sitting, standing, breaks in sitting, steps, sleep duration, sleep quality | Worn in a belt around a hip, or a clip attached to trousers, worn in a wrist during nighttime | Bluetooth, 3D accelerometer, gyroscope | None. Data access through a mobile app (Exsed2). The app visualizes the recorded data on a daily graph and a weekly graph. | Yes, up to 30 meters. | M = 83% of the time (19 h 55 min /day) |
| Firstbeat Bodyguard 2 | Heart rate variability (HRV) sensor | Stress, recovery, duration of physical activity with intensities, HRV, heart rate, EPOC, respiration rate, others | The device is attached to the chest with two disposable clinical grade ECG electrodes. | 3D accelerometer, beat-to-beat heart rate | None. Data is provided in a PDF after the measurement period. | No. | M = 93% of the time (22 h 30 min / day) |

**Textbox 2.** Main interview questions regarding the wearable sensors

> **Main question about the self-discovery**
> - Have you made deductions based on the data from the sensors and their apps? If yes, what kind of?
> - Has the usage of the sensors influenced your behavior? If yes, how?
> - Do you think that the <sensor name> would help you to manage blood glucose? Please justify.
> - Has the information from the sensors or their apps been confusing or unclear? If yes, what?
> - Did you feel that the information from the sensors described your behavior truthfully?
>
> **Main questions about the user experience**
> - What factors influenced wearing the sensors?
> - Have the sensors or their apps caused you any discomfort or inconvenience? If yes, which sensors or apps and how?
> - Think about your experience with the sensors and their apps. How would you improve them?

# 3 RESULTS

## 3.1 Participants

Ten women with GDM (see Table 2) were recruited. We had a variety of participants in terms of age (min 24 years, max 40 years). Participants were familiar with mobile apps and measuring glucose, but they had less experience with using wearable physical activity sensors as depicted in Table 2. The same participants participated in another study after this study [23]. The mean age of the participants was 33.6 years, which is similar to women with GDM in Finland (32.5±5.3 years) and in Helsinki area (33.1±5.1 years) [77]. The mean body mass index (BMI) of the participants was 25.7 kg/m$^2$, which is in the range of the mean BMI of women with GDM in Helsinki area (27.1±6.0 kg/m$^2$) and in Finland (28.5±6.3 kg/m$^2$) [77].

**Table 2.** Participant demographics and their experience with mobile apps and sensors. Regarding the statements, the Likert-scale was from 1 (=Strongly disagree) to 5 (= Strongly agree).

| ID | Age | Weeks of gestation | Body mass index before pregnancy (kg/m$^2$) | How many minutes per day do you exercise at a moderate level? | I am used to use various mobile apps | I am used to using physical activity sensors (such as Fitbit, Vivosmar and Polar | I am familiar with measuring blood glucose. |
|---|---|---|---|---|---|---|---|
| 1 | 36 | 35 | 22,2 | 150 | 4 | 3 | 5 |
| 2 | 32 | 33,3 | 30,1 | 4 | 4 | 2 | 4 |
| 3 | 40 | 31,2 | 23,1 | 120 | 4 | 2 | 4 |
| 4 | 24 | 33,7 | 29,8 | 240 | 5 | 2 | 5 |
| 5 | 31 | 35,6 | 26 | 3 | 4 | 2 | 4 |
| 6 | 31 | 30,3 | 21 | 210 | 5 | 1 | 4 |
| 7 | 32 | 36,6 | 20,2 | 3 | 2 | 1 | 5 |
| 8 | 36 | 37 | 25,4 | 120 | 5 | 5 | 5 |
| 9 | 35 | 34,8 | 22,9 | 120 | 5 | 1 | 4 |
| 10 | 39 | 28,1 | 36,6 | 150 | 5 | 5 | 5 |
| average | 33,6 | 33,6 | 25,7 | 25,7 | 4,3 | 2,4 | 4,5 |

## 3.2 Factors supporting Self-Discovery (RQ1)

### 3.2.1 Continuous Glucose Monitoring

While participants were familiar with measuring their glucose levels (see Table 2), they learned new things due to continuous monitoring (P9: '*(I wish I had this [CGM] when I got the GDM diagnosis, so I would have got some knowledge of the glucose curve.*"). They learned new causalities between food and glucose levels (P1: "*I think it is better to have the data from 24 hours. Then you can see what happens in between. Nowadays, I eat nuts because I know that when I started eating nuts, my blood glucose started to be at a good level.*"). Another example of improved glucose control was with P2, who started following her glucose continuously and learned to adjust her eating accordingly: P2: "*I had a couple of hypers [hyperglycemia], but I think with normal measurements those would not be noticed because they were irregular...especially the hypers in the morning...At first, I was like I don't have any problem with them [glucose levels] but when you had that continuous measurement I figured out that it is not actually the case.*". So, the CGM facilitated following the variability of glucose, and for seven (7/10) participants the variability of their glucose values was decreased, calculated as a trend in the variability of glucose using LAGE (Large amplitude of glucose excursions) [78]. CGM not only supported self-discovery but also improved motivation to change diet (P2: "*...you are able to see it [glucose] for the whole day...it motivates for changing the diet.*"). While participants had extra costs from wearing the CGM (see Section 3.4.1) and calibration (See Section 3.4.3), most of the participants would have liked to continue using the CGM, as they got used to it.

### 3.2.2 Numerical Affirmation for Assumed Cause-and-effects

Half of the participants discussed that they found the numerical evidence for the assumptions they had before the study (P10: "These *sensors have confirmed my assumptions what are the most important factors to control blood glucose and GDM and weight management in the future...so the regular eating is of paramount importance for me.*". Also, this included more specific causalities they had assumed before using the sensors, as P9 found evidence for the association between physical activity and blood glucose (P9: " *If you move or plan to move, then you can eat food which has more carbs...so I have been following if I do something I can eat a little bit more...this kind of normal thing that I kind of had thought before...but now it was more like you can actually see it.*").

## 3.3 Factors inhibiting Self-Discovery (RQ1)

Most of the participants (7/10) did not discuss finding cause-and-effects between physical activity and glucose levels. For example, P9, who was data-oriented, tried to figure out the causalities (P9: "*Well, maybe the information from the activity bracelet was useful, as I have never used such a device before and I am interested in numbers... and this information connected to what is happening in my blood glucose...so I tried to figure out connections.*"). As the self-discovery process seemed to be tedious for many of the participants, they would have liked to receive clear instructions on how to change their behavior. Some participants wished to have seen important data popping-out (P10: "*I wished I could have seen highlighting or other markings, what to look for from the data*"). As such, the current tools did not support establishing links between glucose levels and physical activity. In the following, we discuss issues that inhibited self-discovery.

### 3.3.1 The Lack of Trackable Features

Participants had less physical activity than recommended during the measurement period as measured with Firstbeat. According to recommendation, pregnant women should have at least 150 mins of moderate physical activity in a week [79], but according to Firstbeat the participants had approx. 7 mins/day (see Figure 2). In most cases, the lack of physical activity was explained by being at the third semester of pregnancy (P2: "*Unfortunately, I did not have much physical activity as I get pain from normal walking...I was tempted to do more, but my condition did not allow it.*"). Thus, without enough physical activity, it is difficult to interpret its effect on glucose.

As the intensity levels of physical activity were difficult to quantify and recognize, the participants had only very little understanding of what the physical activity shown as intensity minutes meant (P4: "*They were very confusing, I did not follow them actively, one day I just realized that I have got more of them, but I did not have any clue what they are based on. On one day I became unwell in a shop, and I noticed that I had received intensity minutes because my heart rate had increased...but it was not something nice.*"). The number of intensity minutes varied a lot between participants, as one participant did not gain intensity minutes at all during the measurement period and one participant gained 145 mins (the goal being 150 mins per week). Moreover, Vivosmart 3 required physical activity to last 10 consecutive minutes to be counted, which was not often the case for participants as their physical activities consisted of shorter periods such as walking the stairs.

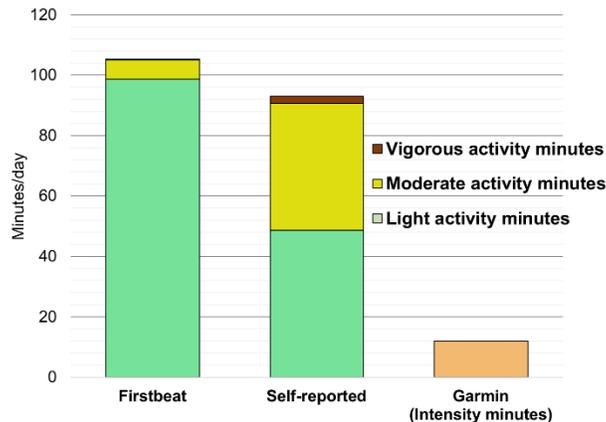

Fig. 2. Duration and intensity of daily physical activity as measured with Firstbeat (HRV) and self-reported. A substantial portion of physical activity that HRV-measured to be light was perceived as moderate.

While the intensity of physical activity was difficult to recognize and intensity minutes were not achieved much or understood well, steps were easily understood, and step goals provided by sensor applications were achieved more often. However, half of the participants did not care about the goal, as walking was perceived to be tedious (P5: "*I did not care about the step goals, before pregnancy I could have challenged myself, but now I go for a walk which feels good and that's it.*"). Three participants (3/10) discussed the importance of the possibility to track swimming and water running, as they were the only exercises they were able to do well (P1: "*For a gestational diabetes patient, swimming is almost the only sport that you can pretty normally do, so the sensor should definitely be one that encourages you to move, especially to swim.*"). This highlights the importance of waterproofness of physical activity trackers and the possibility of track swimming for women with GDM.

3.3.2    *Difficulty of Quantification of Self-tracking Data*

We expected that quantified information through wearable sensors would have helped in forming hypotheses, as an abstraction to quantifiable units (e.g., from a fast walk to heartbeat) is often required at the hypothesis formulation stage [24]. However, the discrepancy between perceived and measured quantification and clearly erroneous quantification with wearable sensors imposed significant challenges for hypothesis formulation. This study showed a significant difference between measured and perceived quantification of physical activity. Participants interpreted the intensity of physical activity higher than it was measured, that is participants perceived light activity as moderate. This can be seen in Figure 2, which shows the high portion of physical activity being light as measured with Firstbeat. The participants self-reported their overall duration of physical activity rather similarly with Firstbeat. In fact, there was a statistically significant correlation between Firstbeat and self-reports (Spearman r(59) = .43, p < .01) on the duration of physical activity. However, the participants categorized the intensity (intensities were instructed according to [69] of the physical activity differently than Firstbeat. There were no statistically significant correlations (p > .05) between self-reported values and the values from Firstbeat when looking at each intensity within the categories.

In general, the participants had difficulties in interpreting what is counted as physical activity: (P4: "At this point of pregnancy you move a little, and tasks like fetching the mail is already pretty tough...so it is a bit difficult to say what is counted as exercising and what is not.)". As such perceiving the physical activity as more intense than measured might lead to misconclusions about its effect on glucose levels.

### 3.3.3 Contradicting Self-tracking Data

The differences in the data provided by the sensors induced significant challenges for self-discovery. Regarding physical activity, there were statistically significant differences in the number of steps between the devices, as evaluated with the Friedman test, $\chi_2(2) = 16.22$, p < .01. The differences were not only explained by the differences in how long the sensors were worn, as Vivosmart 3 and Firstbeat were worn similar amounts, but Vivosmart 3 (M = 7191 steps/day) provided twice as many steps as Firstbeat (M = 3519 steps/day). Exsed was in the middle with M = 6307 steps/day. Firstbeat required a longer continued movement to start the counter, whereas Vivosmart 3 started counting the steps immediately. It is probably a more desirable strategy to also count the steps during small transitions e.g., in the home, as there were only a few pregnant women who exercised. However, Vivosmart 3 counted movement as steps, even though the participants had not walked (P1: "*when I woke in the morning I had several hundred steps, although I had not walked that much during the night.*"). Contradicting data between the sensors was not only limited to steps, as there was not a significant correlation in duration of moderate physical activity between Firstbeat and the amount of intensity minutes in Vivosmart 3 (Spearman r(57) = .22, p = .12).

Regarding sleep, there was a statistically significant difference in the length of sleep between the devices, as evaluated with the Friedman test, $\chi_2(2) = 17.27$, p < .001. The Exsed showed significantly less sleep (M = 7.2 h/night), compared with Vivosmart 3 (M = 7.8 h/night), and Firstbeat (M = 8.0 h/night). These differences raised a lot of questions among participants and decreased the credibility of the data. These responses on contradicting data also reflect on the UTAUT-responses on incompatibility (see Section 3.4.4). Three participants found that the data provided by the sensors they normally use (activity bracelets by Fitbit, Polar, and Suunto), varied significantly in terms of physical activity and sleep.

In addition, six participants (6/10) discussed differences between continuous glucose measurements taken from tissue and fingerstick measurements taken from blood. The reported differences varied a lot, as some reported they were significant (P4: "*...a couple of times it [Medtronic] showed that the glucose was low, but it wasn't that low... at one time it [Medtronic] showed 2.8 [mmol/l], but it was 5.3 [mmol/l, as measured from fingertip].*"), and some reported they were minor (P6: "*I don't think they differed much...looking at the graph you were able to see an increase after eating and during night time it was low, so they seemed to be pretty*

*accurate.*"). Nevertheless, the differences decreased the credibility of glucose-monitoring data (P9: "*...the values were somewhat different than taken from fingertip...so it made me think how much I can trust this data.*"). However, the use of multiple sensors supported gathering a lot of data from many perspectives that have the potential to increase understanding.

### 3.3.4 Challenges in Self-Tracking of Sleep

As pregnancy decreases the quality and length of sleep [80], sleep information could be valuable for women with GDM, as they learn to understand their sleeping disorders. Five participants (5/10) mentioned information about sleeping to be particularly interesting (P10: "*On Thursday night I slept two hours and six minutes, so it was pretty interesting to get that kind of readings, but I think it is positive in the sense that it proves that I am not becoming crazy but instead slept too little*"). Also, these participants discussed that they were interested in the quality of sleep (P9: "*It was interesting to look at the sleep graph...so in the early night I had slept deeper and lighter towards the morning, and how you have woken or not woken up.*"). However, two (2/10) participants did not want to get feedback about their sleeping as they knew they had slept too little (P1: "*I have not had any possibilities to influence my sleeping during the past month, so it could be a bit depressing information that you have slept lousy...Well, I know that already.*"). Thus, seeing sleep data was clearly a matter of personal preference.

Participants had sometimes difficulties in estimating at what time they had fallen asleep; thus, the objectively measured sleep has the potential to provide unbiased information for the self-discovery process. In general, participants' self-reported duration of sleep (M = 7.8 h/night) correlated with duration of sleep measured with wearable sensors: Firstbeat (Pearson $r(42) = .58$, $p < .001$), Vivosmart 3 (Pearson $r(42) = .57$, $p < .001$), Exsed (Pearson $r(36) = .55$, $p < .01$). Moreover, sleep data gathered through sensors is more comprehensive as participants sometimes forgot to mark the waking and sleeping times in the logbook.

Nevertheless, participants were not able to link their sleep with glucose values, although they tried to increase their understanding of how to manage glucose values (P8: "*I am most interested in the quality of sleep and stress levels. And how and if they impact the glucose somehow...my fasting glucose values don't seem to be within the limits no matter what, so it is the same whatever I eat, so I feel that they are always high.*").

### 3.3.5 Challenges in Self-tracking of Stress and Recovery

In general, all the participants were curious about their stress levels and how it was linked to glucose levels. However, most of them (7/10) had difficulties in interpreting the stress data provided by Vivosmart 3. Pregnancy increases the resting heart rate and decreases the HRV [81] which has been used as a measurement for stress[82]. The decrease in HRV due to pregnancy most likely caused Vivosmart 3 to interpret the standing as stress, although the participant did not feel stressed (P3: "*The stress data was confusing. I did not understand how it figured out that I had been very stressed that day. I stood a lot at my workstation, so I wondered if it is so silly that it thinks that I am terribly stressed if I stand.*"). However, three participants (3/10) valued the stress data from Vivosmart 3 as it helped them know whether they have recovered from stress (P9: "*There was one day when I was using a computer and I had meetings for the whole day, it was very stressful for the body, even though I did not do anything physically...these stress sensors sort of gave me information on what is enough rest for recovery, this was new to me.*"). Seeing themselves being described as stressed did not seem to make them more stressed but helped sometimes to distinguish between stress and rest (P7: "*I was able to look at the stress level, so it concretized when I am like resting and when the stress is high.*"). P8 discussed that stress reading from the sensor could be used as an objective value like body temperature, which would make the partner understand their condition (P8: "*...at home I can show, look how stressed I am...so you*

*should take care of the child while I'm resting."*). Thus, stress data was valued by other means than supporting the self-discovery of glucose levels.

### 3.3.6 Towards Better Tools Supporting Self-Discovery

Although participants had received their GDM diagnosis some weeks prior to the study's measurement period, they were still in the discovery phase [38], meaning that they were figuring out the factors affecting their glucose levels. We found many instances that followed the chosen self-discovery framework [24]. Over half of the participants (7/10) found causalities between nutrition and glucose values in CGM and three participants (3/10) between physical activity and glucose values in CGM. However, these causalities were based on the gained experience (i.e., the food she just ate or walk she just went for) and the CGM data, but not on the data by the lifestyle sensors. This indicates that establishing the causalities based on self-tracking data through wearable sensors seemed to be too challenging and better tools (or more support from health care professionals) for interpreting the self-tracking data through wearable sensors are needed. In this study, six (6/10) participants commented that they would have valued if they could have added information to one single app, which also would have decreased the amount of redundant data shown (P10: "*So that the same information would not be entered in many places, but also the same or overlapping information would not be presented to the user, so you should have one app.*"). This was reflected by P9 (P9: "*So that all the information is visible in one place, and there won't be many links and sources. So, the challenging thing was what I should write on the paper, what I see on the bracelet...so there should be one place and one way to show this information.*"). The other issue was that participants had to enter the blood glucose values taken from their fingertip into the Medtronic app and write them down with pen and paper and report these values to a health care professional. This required double marking of blood glucose values, which has decreased the motivation to track the glucose values in the GDM application in the long term [42]. As such, participants indicated that they wish to have a single application where all the data from lifestyle sensors and the CGM is gathered. This would decrease the amount of redundant and contradicting data as now participants were confused by the differences in the data provided by multiple sensors.

## 3.4 Experiences of Wearing the Sensors (RQ2)

### 3.4.1 Wearing the CGM on the Arm

Most of the participants (8/10) preferred wearing the CGM on the arm instead of near the navel. The reasons were that participants did not like to attach the sensor to near the baby (P6: "*Now when you feel with your hands your baby moving, it would feel somehow weird if there was something in that place during pregnancy.*"), the abdomen was sore, and that the sensor would be visible to self and others. However, wearing the CGM on the arm caused problems for glucose measurement during night times, as the participants slept on the glucose sensor which lowered the sensor readings to go below the limit of alarm and this woke up most of the participants (8/10). The participants had to turn off the iPod to silence the alarm which caused some of the glucose measurements to be missing from the sensor. So, the participants could not sleep on the side where the sensor was placed. We tried to avoid this by asking on which side the participant typically sleeps and attaching the sensor to the other side, but this did not always help as some participants sleep on both sides (P1: "*At this stage of pregnancy...you must sleep on both sides, they are the only poses in which you can sleep, so the position has to be something else than that [the arm]...*").

While most preferred not to wear the glucose sensor near the navel, P10 would have preferred that option. She hit the glucose sensor on various places such as a car seat (P10: "*For example, I hit it [glucose sensor] on the car seat every time I got in the car or got out of the car it hurt...so I wonder if there is a better place for it.*"), and in fact, three (3/10) participants reported the issues of hitting the sensor on various objects causing some pain in the arm. As such, there was no optimal place where this CGM could be placed. The other issue with stickers is that they can be loosened when swimming, which was an important hobby for the three participants. In fact, the stickers holding the CGM were loosened for one participant and the sensor got detached when swimming. Therefore, stickers as a fastening mechanism in sensors should be avoided in the long run (P1: "*...six days is pretty heavy, so you do not want to take them all with you, so I think, especially when there are these glues, so I would not like to wear them for very long.*").

### 3.4.2 Wearing the Lifestyle Sensors

Overall, the participants wore the sensors over 80% of the time (i.e. over 19 h/day), as shown in Table 1. Participants wore the sensors, except when they were showering or when swimming. Sometimes they forgot to wear the sensor, and this was especially the case with Exsed (hip worn) that required a change of position before and after sleeping. In fact, there was a statistically significant difference in measurement durations, as evaluated with the Friedman test, $\chi_2(3) = 8.124$, $p = .044$. A post-hoc test using Bonferroni correction [83] revealed that data from Exsed was acquired for a statistically ($p = .03$) significantly shorter time (M = 83% of the time) than from the Vivosmart 3 activity bracelet (M = 93% of the time). No other differences were found between them in terms of measurement durations. The Exsed hip sensor operated with batteries during the whole period and did not require charging, however, the Vivosmart 3 was worn more. Participants had only little wearing issues with Vivosmart 3. Two participants discussed that it caused some swelling, but no other issues were raised.

This was different from Exsed worn on a hip. Four (4/9 - we forgot to ask this from one participant) participants preferred to wear the Exsed in a clip, and two (2/9) with a belt, and three (3/9) did not have a preference. The primary reason participants preferred wearing the Exsed on a clip was that it was difficult to adjust the tightness of the belt. When the belt was loose, it easily moved around (P10: "*It rolled all the time and fell down, so it was a bit irritating.*") or if it was tight, it pressed uncomfortably (P8: "*The belt pressed even more [than the clip], I do not know how much it could have been looser.*"). The belt was used if no place was available for the clip (P5: "I *am wearing a skirt or dress, so the belt has been more natural.*"). Participants had issues with the clip, as it chafed the skin (P2: "*As I have this belly, it [Exsed] is irritating on the waist. ...I had to fix its [Exsed] position and move it so if I am sitting it is under pressure.*"). In fact, five (5/10) of the participants reported that they had some issues with wearing Exsed with either clip or belt. As such, the pregnancy decreased the feasibility of using a hip sensor for tracking physical activity. However, the hip sensor was perceived as unnoticeable by some participants as it did not have a user interface and it was worn in the trousers (P4: "*You did not notice it at all, so sometimes I forgot that I needed to put it on when I took my trousers off*.").

### 3.4.3 Needed Effort Using the Sensors

The requirement to calibrate the CGM twice a day was found to be tedious (P2: "It [calibration] was needed surprisingly often...although it did not bother me during the week, but in the long term it could become an issue, all those calibrations if you are somewhere [else than home]..."). This influenced P4's sleep, as she needed to wake up in the mornings to calibrate (P4: "On some mornings, it was irritating that it notified me half an hour before calibration, I thought I could have slept half an hour more."). The other issues that needed substantial effort from participants, was keeping the nutrition diary and filling the physical activity logbooks. These would not be feasible in the long-term (P3: "Writing the diaries took a lot of time. I could not manage that every day."). These responses support the findings from [42] that the requirement of manually entering physical activity reduces the amount of data significantly in the long run among women with GDM. Even manual start/stop for recording exercises was not used much, as it was easily forgotten (P10: "...it was very difficult to remember to mark the activities, like starting the activity and stopping the activity.").

*3.4.4  General Acceptance based on UTAUT*

Responses to the UTAUT-questionnaire (see results in Multimedia Appendix 1) showed good acceptance of sensors before and after usage. For example, participants agreed with the statement "*I would find using the sensors as a good idea*" (Before M = 6.0, After M = 6.1 out of 7 where 7 is "Strongly agree"). Participants felt that wearable sensors supported behavior change, they agreed with the statement "*Using the sensors will improve my possibility to make a concrete improvement in my lifestyle*" (Before M = 6.0, After M = 5.9 out of 7 where 7 is 'Strongly agree'). Participants mentioned that being able to see trends could guide their behavior related to diet and physical activity.

The acceptance was not affected by the usage of the sensors, as there was no statistically significant difference ($p > .05$) between acceptance before or after the usage (evaluated with Wilcox signed-rank test). The largest difference between before and after the usage was in the statement: "*The sensors are not compatible with the other sensors I use for self-tracking*". Before the study participants disagreed with the statement (M = 2.5), but after they slightly agreed (M = 4.5). Only the participants who were using other self-tracking sensors responded to this statement, so the sample size was too small to conduct a meaningful statistical test. However, the responses in the interviews reflected the change in responses on incompatibility (P10: "*I found differences in both activity sensors [Exsed and Vivosmart 3] compared to this my own Polar, which was on my other hand. I changed its settings to correspond with the right arm...it [Polar] gave different readings on activity and steps, although the length of a step was set to the same. It was so mysterious why they differed so much.*"). Despite this incompatibility with the participants' existing self-tracking devices, the use of wearable lifestyle sensors together with CGM was acceptable.

## 4  DISCUSSION

This is the first study which aimed to study how to support self-management of GDM with wearable sensors in addition to CGMs. Regarding self-discovery (RQ1) we found that the CGM supported the learning of the associations between blood glucose and nutrition, but the wearable sensors measuring physical activity, sleep and stress did not provide significant support for the learning. The challenges included dispersion of data between multiple apps, missing trackable features, such as type and intensity of physical activity, and the lack of GDM specific goals for behavior. From the user experience perspective (RQ2) this study highlighted that the benefits overcame the discomfort and effort wearing the sensors. There were differences between preferences with sensors, and wrist-worn sensor was preferred over hip-worn sensor and was worn more. In general, this study further emphasizes the findings [22,43] that self-tracking among women with GDM should be highly automatic. We discuss these results in following sections with respect to each RQ.

## 4.1  Supporting Self-discovery with Wearable Sensors (RQ1)

*4.1.1  Feature selection*

Starting from *feature selection* (i.e., identification of activities that have impact on blood glucose), this study highlighted the need to tailor the available features and their presentation with respect to GDM. Women with GDM had difficulties in interpreting and accessing the physical activity features. The activity bracelet required users to have physical activity at a moderate level for 10 consecutive minutes to get the duration of physical activity visible, which was not often the case for the women with GDM as they did small activities, such as short walks. In fact, two participants did not achieve intensity minutes at all. This also would mean showing light physical activity, for example in terms of steps. However, there are no official health recommendations for steps for pregnant women, and thus, showing the duration of moderate or vigorous physical activity with respect to health recommendation (150 mins/week of physical activity at a moderate level [79]) would be a feasible feature on a weekly basis.

Although we used multiple distinct types of wearable sensors for measuring physical activity, there was a lack of automatically recognizing physical activities (i.e., swimming and water running) that are important for women with GDM. This challenge will decrease in the future as the automatic recognition of diverse types of physical activities is improving. However, this challenge of automatic recognition of features related to nutrition will remain for long. To cover a wide variety of features, MacLeod et al. [27] suggested the use of manual tracking as an aid to automatic tracking. This approach allows tracking a large number of features. However, qualitative studies emphasize that pregnant women are typically overwhelmed [54,84] and that women with GDM face considerable time pressures [84]. As such, we argue that automatic self-tracking is especially important for these user groups. In this study, most efforts were requested for keeping a food diary with pen and paper. However, less demanding methods for this were requested. Chung et al. [85] proposed a lightweight photo-based food diary to support the collection of nutrition data for clinical visits of patients with irritable bowel syndrome. This photo-based diary approach appears to be promising for women with GDM as well. Peyton et al. [54] suggest that self-monitoring of pregnant women can be supported and encouraged, in addition to photographic journals, by using simple designs, such as reminders, and by keeping the techniques for user data input simple. Data collection techniques that are undemanding (e.g., checkboxes instead of long texts) support quantifiable format, which is needed in the *hypotheses formulation* process [24].

### 4.1.2 Hypothesis formulation

With respect to *hypothesis formulation* (i.e., formulation of suspected associations with activities and blood glucose) participants experienced difficulties in quantification of the self-tracking data on physical activity, sleep and stress. Still, most of the participants were interested in following stress which plays a significant role in the lives of women with GDM [13,86] and sleep to follow sleeping disorders due to pregnancy. Thus, this quantified information about stress and sleep provided value for the participants in terms of providing information about their condition, being part of *documentary tracking* [49]. As such, participants were interested in following their sleep and stress rather than changing them. This was opposite to nutrition and physical activity, which were more related to *goal-driven tracking* [49], and their features (although not based on self-tracking data) were an integral part of the self-discovery process.

The results of this study indicate that quantification by the sensor needs to match with quantification by the user so that meaningful hypotheses can be formulated. For physical activity, misperception of intensity is problematic as the rate of change of glucose levels depends on the intensity of physical activity [87] and perceiving the physical activity differently may lead to wrong conclusions about its effect. This finding of the discrepancy in perceived and measured intensity of physical activity is in a line with [88], where women with GDM estimated the amount of vigorous physical activity higher than measured with hip worn accelerometer. These results are opposite to the results of a study [32], where users with type-2-diabetes reported a high correlation between self-reported physical activity and duration of vigorous activity measured with an activity bracelet. This indicates that the discrepancy between perceived and measured physical activity is more prominent among pregnant women than with people with type-2-diabetes. For women with GDM this would mean that the intensity levels should be more clearly defined for women with GDM, and providing feedback during the activity (e.g., "Now you are swimming at the moderate level.") would be a good approach. Moreover, the quantification of features with wearable sensors was unreliable, for example, participants could not rely on stress data which was affected by decreased HRV due to pregnancy. Thus, we agree that more advanced techniques are required to differentiate between the decreased HRV caused by pregnancy and decreased HRV due to stress [59].

*4.1.3    Hypothesis evaluation*

For *hypothesis evaluation* (i.e., evaluation of how latest information about associations fit with existing knowledge), we observed the challenges of scattered and conflicting Data. At this stage, we expected that having the wearable sensors would have facilitated hypothesis evaluation, as there is more data available, and its quantified form enables quantitative comparison against existing data. However, we found two major challenges why this stage was difficult for the participants. First, the data was scattered to different apps, making comparisons between lifestyle and glucose tedious. The dispersion of data has been identified as a challenge in personal informatics [26,89,90] and this study further emphasized that there should be integrative tools to support self-discovery.

The data was contradictory between sensors in multiple ways, for example, there was a statistically significant difference in the number of steps and duration of sleep between the sensors. Moreover, the discrepancies between CGM and fingerstick measurements confused how much participants could rely on CGM data. The discrepancies in data were not only limited to given sensors but also with data from participants' existing sensors (see Section 3.4.4). These discrepancies directed the attention of women with GDM from self-discovery to evaluate these differences. While using multiple sensors potentially increases the reliability of the data, the use of a single sensor for each shown lifestyle variable would be more appropriate to support reflection. Then the attention of the user is not on looking at differences in data between sensors, but to evaluate the impact of activities on glucose levels between instances, such as small variations in meals and physical activities. Then the relative differences in data within one wearable sensor would provide useful information. However, we acknowledge that trackable features may be unknown for persons with chronic illness, especially in *poorly-understood conditions* [27]. Then figuring out the relevant features may require the use of multiple wearable sensors to gather various aspects of chronic illness. However, in that case, the data from multiple sensors should not be conflicting, but supportive for increasing the understanding of the chronic condition.

*4.1.4    Goal setting*

On a high level, the goal for behavior women with GDM is simple. The fasting glucose value in the mornings should be less than 5.5 mmol/l and the glucose value one hour after a meal should be under 7.8 mmol/l. However, this is a very high-level goal, which participants try to transform into concrete behavioral goals. For the *goal specification* (i.e., identification of future goals based between activities and outcome) phase in self-discovery, we found that participants primarily created goals based on CGM and experience. Of all the target behaviors, changing diet was the one the participants seemed to be most optimistic about, and they could name several ways how they changed it.  For example, P1 defined a goal of eating nuts in the meals as she figured out that helps to keep her glucose under the maximum limit. To help in goal setting, the woman with GDM should know how many nuts or how many grams of nuts to include in the meals, and she should have a tool to track this goal, developed following goal-directed self-tracking approach [91]. Transformation of goals defined by the participants (e.g., eating more nuts by P1) into features, which are possible to track with wearable sensors is still a major challenge.

Goals provided by wearable sensors (e.g., 150 mins/week of physical activity at a moderate or vigorous level) were related to general guidelines, but not specific to the management of GDM. This decreased their value for women with GDM. Some limitations are part of every chronic illness, and individuals with chronic illness should not be pushed too hard to achieve the goals, as there is a risk of causing *goal frustration* (cf. [92]) if it is impossible to achieve them due to implications from their illness. The goals should be concrete (e.g., "Walking for 30 min at the moderate level would decrease your glucose levels") and trackable with wearable sensors. The other type of goal specification we observed was that participants defined goals to collect further evidence for their hypothesis. For example, for P3 the goal was to climb stairs to see whether this had a real impact on her glucose levels. Again, this goal should be trackable. Half of the participants discussed that they would be willing to change behaviors for physical activity. One reason was that physical activity was measured in a straightforward way (i.e., steps) and was experienced as more tangible by the participants than the target behaviors related to sleep and recovery (see Sections 3.3.4 and 3.3.5).

One way to approach this is to provide options for concrete goals, where the women with GDM could choose the most preferred one(s). Having such a set of options for goals would ease the tracking with wearable sensors, as the number of trackable features and goals could be narrowed down to certain options. Harrison et al. [93] suggested having practical options for goals for encouraging physical activity for women with GDM, as Harrison et al. [93] found that women with GDM wish to have clear goals for physical activity while still retaining autonomy. We made similar observation for nutrition goals. The requested goals did not only include what to eat considering the diet limitations (e.g., due to pregnancy), but also when to eat. This reflects the wish of people with type-2-diabetes who have experience with CGM to have more knowledge on the effect of meals on temporal glucose patterns [94]. While we made the same observation, the women with GDM wished to have concrete suggestions on how to influence these glucose patterns by content and timing of the meals.

## 4.2 Experiences of Self-Tracking with Wearable Sensors (RQ2)

We learned that the body placement of sensors is a key factor in acceptability, quality of measurements, preference, and ultimately a challenge for collecting data. Wearing the physical activity sensor on the wrist, instead of on the hip, has several benefits for pregnant women. Half of the participants had issues with wearing the sensor on their hip, as it moved around and/or chafed the skin when sitting. The drawback with a wrist-worn sensor is that it is not possible to recognize whether the user is sitting or standing. A sensor worn on a hip can recognize this [63]. However, regulation of sitting and standing relates more to long-term health and health risks [95] rather than to the management of GDM. The sensor on a wrist was worn statistically significantly more than on a hip, providing more data to the user. Although the hip-worn sensor was used less than the activity bracelet, it was still worn for more than 10 hours in a day, which has been the minimal amount to get credible data [96].

Wrist-worn sensors are particularly feasible for pregnant women as bracelets can be adjusted with respect to swelling. This is not the case with activity rings, such as Oura, which are not easily worn during pregnancy due to swelling of fingers [97]. While there are wearability issues with wrist-worn devices among pregnant women, such as smartwatches if they are heavy [97], the activity bracelet used in the study did not raise issues beyond slight irritation of the skin. This finding has evidence from a long-term study conducted with a similar activity bracelet among pregnant women [98].

The lifestyle sensors were highly accepted among women with GDM. This result extends the finding by Scott et al. [46] that CGMs are highly accepted in self-tracking during pregnancy. Women with GDM seem to be less concerned about using wearable sensors compared with chronic illnesses, such as chronic heart patients who have had feelings of uncertainty, fear, and anxiety [99]. In our case with GDM patients, the clear purpose of the wearable sensors (supporting self-discovery and healthy behavior) could have increased the acceptability of sensors. This was the opposite in the case of heart patients where the purpose of sensors was to gather "self-tracking of activity data in relation to their embodied condition and daily practices of dealing with a chronic heart condition" [99]. Thus, we expect the framing of the purpose of wearable sensors clearly and supporting the goal of the user (in this case to manage glucose levels) with wearable sensors seem to increase the acceptability of self-tracking.

Although the data provided by CGM was highly valued among participants, most of the participants had issues wearing the CGM. Most of the participants preferred wearing the CGM on the arm, instead of having it near the navel, which is the primary placement for the sensor. Wearing the sensor on the arm caused false alarms of glucose going too low because women with GDM slept on top of the sensor. As such, if CGM is worn on the arm, a more robust sensor against pressure is needed as pregnant women tend to sleep on their side, at least when over 30 weeks into gestation [100]. Moreover, due to placement on the arm, participants could not attach the sensor themselves. This decreases the feasibility of using this CGM in the long term, as the CGM needs to be recharged once a week and the sensor can detach, for example, due to swimming (see Section 3.4.2).

To support self-management having a single "output" i.e., a GDM application where all the collected data would be shown in a single view (see Section 3.3.6) also induces a question of having a single "input" i.e., a wearable sensor that collects all the data. A feasible approach would be adding lifestyle tracking capabilities to CGM. This kind of sensor does not exist yet. An integrated sensor would decrease the problems of wearing and managing multiple sensors, and the data would be recorded in synchrony and without discrepancies and, thus, helping in establishing the causalities between lifestyle and glucose levels. Moreover, having a single sensor would remove the technical work required to integrate data from multiple cloud services [101]. Ultimately, this integrated sensor would be worn on a wrist. Having a wrist-worn sensor would remove difficulties that were raised from the CGM worn in behind the arm (especially causing false alarm during nights) and from physical activity sensor worn in the hip. However, non-invasive glucose tracking from the wrist has suffered from poor accuracy resulting from movement, exercising, and sweating [102]. Thus, the optimal solution for one single wearable sensor is still to be developed.

While we focused on self-discovery without the help of health care professionals, they were very often mentioned. The continuous data collected by the wearable sensors provides an opportunity for remote monitoring and feedback by health care professionals [60]. The participants discussed the importance of having contact with a diabetes nurse, so that they can share the data with them and discuss the data provided by the glucose sensor. This is in line with previous findings, that persons with chronic illness need help from experts in the self-discovery process [24,27] and behavior change. This is further supported by reviews on technological support for diabetes management, which emphasize the importance of two-way communication between people with diabetes and health professionals [103,104]. Further, self-tracking with wearable sensors can increase the completeness of the self-tracking data presented to health care professionals [105] and can increase the perceived usefulness of the sensors [103,104]. Thus, at this stage, having a two-way channel between women with GDM and diabetes nurses (e.g., through a text chat as suggested by one participant), would still be a crucial factor in supporting the management of GDM.

Although no other wearable sensor than CGM supported self-discovery, they increased self-awareness of one's own lifestyle and women with GDM believed that this would help them to improve their habits. Thus, wearable sensors have the potential to support behavior change for women with GDM, as self-tracking itself has been found to be an effective behavior technique amongst people with type-2-diabetes [52]. However, participants discussed that behavior change should be facilitated with recommendations, which would be formulated either automatically based on self-tracking data or manually by health care professionals. and further, the use of artificial intelligence approaches for increasing the understanding of cause-and-effect relationships [55,106]. This understanding can be used for setting personal goals for lifestyle changes for women with GDM [107], which were highly requested by participants of this study.

### 4.3 Study Limitations and Future Research

We acknowledge that the number of participants could have been higher. However, the main approach of this study was qualitative, and we believe that the number of participants was enough as no new codes emerged after eight interviews indicating the saturation of the data. Moreover, the same number of participants have been used in qualitative studies on experiences of GDM (e.g., [14]). The quantitative investigations on the acceptance of self-tracking among women with GDM would require a longer usage period with more participants.

Women with GDM wore multiple wearable sensors at the same time in this study, which might have affected their acceptance. Despite this, responses to UTAUT-questionnaire in this study reflected high acceptance of wearable sensors. The high acceptance could have been affected by the fact that the participants volunteered for the study, and thus, showed at least some interest in self-tracking and were not afraid of pricking their skin. In fact, one participant did not want to participate as she heard that the study involves skin-pricking. Therefore, the acceptability could be biased similarly to studies investigating the acceptability of CGMs among women with GDM [44–47].

The self-discovery process of GDM is challenging and demanding, which currently takes a considerable amount of time. Carolan-Olah et al. [84] investigated how the teaching of GDM could be improved, particularly for women with multiethnic and low socioeconomic backgrounds. Cultural differences may pose a need for different trackable features for GDM, for example, water activities among women (e.g., swimming and water running) are less feasible in some cultures [108].

This study focused on CGM and wearable physical activity sensors. As nutrition is important factor in management of GDM, the future work should investigate the use of wearable sensors for nutrition tracking. At the current stage, they are not able to detect the intake of macronutrients (for example carbohydrates)[70,71,109], and thus, their support for self-discovery can be expected to be limited. However, the research on wearable and nutrition collection methods is very active and should be considered in the future.

We have designed a mobile app according to the results of this study and we will conduct a long-term clinical evaluation in randomized controlled trial to explore the effect of comprehensive self-tracking with a mobile app on glucose levels [110].

## 5 Conclusions

We have shown results of a user-centered design process of mobile health intervention for supporting self-management of GDM. Our holistic approach for supporting self-management of GDM with mobile technology included investigations of wearable sensors and a mobile app from self-discovery (learning) and user experience perspectives. We showed multiple issues that inhibit self-management, such as inadequate support for self-tracking physical activity, data discrepancy, and challenges wearing the CGM. One major challenge was the scatteredness of self-tracking data. To support learning further, visualization with guidance through tips and recommendations should be designed to increase women with GDM's ability to manage diabetes in their pregnancy. The design should consider pregnant-specific wearability challenges and requirements for the data gathering and representation proposed in this paper.

### Acknowledgments

We gratefully acknowledge the contribution of Fujitsu Finland Oy for the technical implementation of eMOM app and Elisa for sensor integration. We wish to thank all the women with GDM in Finland taking part in the research, and the study nurses Riikka Lumi, Milla Tuhkanen, and Jaana Palukka. Also, we thank Jari Metsä-Muuronen for providing code for posthoc test in Friedman test. This work was funded by the Business Finland eMOM GDM project.

Conflicts of Interest

None declared.

Multimedia Appendix 1

References

1   Cho NH, Shaw JE, Karuranga S, *et al.* IDF Diabetes Atlas: Global estimates of diabetes prevalence for 2017 and projections for 2045. *Diabetes Res Clin Pract* 2018;**138**:271–81. doi:10.1016/j.diabres.2018.02.023
2   Owens LA, O'Sullivan EP, Kirwan B, *et al.* ATLANTIC DIP: The impact of obesity on pregnancy outcome in glucose-tolerant women. *Diabetes Care* 2010;**33**:577–9. doi:10.2337/dc09-0911
3   Bellamy L, Casas J, Hingorani AD, *et al.* Type 2 diabetes mellitus after gestational diabetes: a systematic review and meta-analysis. *The Lancet* 2009;**373**:1773–9. doi:10.1016/S0140-6736(09)60731-5
4   Gilbert L, Gross J, Lanzi S, *et al.* How diet, physical activity and psychosocial well-being interact in women with gestational diabetes mellitus: An integrative review. *BMC Pregnancy Childbirth* 2019;**19**:1–16. doi:10.1186/s12884-019-2185-y
5   American Diabetes Association. Standards of medical care in diabetes-2016 abridged for primary care providers. *Clinical Diabetes* 2016;**34**:3–21. doi:10.2337/cd16-0067
6   Halse RE, Wallman KE, Newnham JP, *et al.* Home-based exercise training improves capillary glucose profile in women with gestational diabetes. *Med Sci Sports Exerc* 2014;**46**:1702–9. doi:10.1249/MSS.0000000000000302
7   Harrison AL, Shields N, Taylor NF, *et al.* Exercise improves glycaemic control in women diagnosed with gestational diabetes mellitus: a systematic review. *J Physiother* 2016;**62**:188–96. doi:10.1016/j.jphys.2016.08.003
8   Kokic IS, Ivanisevic M, Biolo G, *et al.* Combination of a structured aerobic and resistance exercise improves glycaemic control in pregnant women diagnosed with gestational diabetes mellitus. A randomised controlled trial. *Women and Birth* 2018;**31**:e232–8. doi:10.1016/j.wombi.2017.10.004
9   Onaade O, Maples JM, Rand B, *et al.* Physical activity for blood glucose control in gestational diabetes mellitus: rationale and recommendations for translational behavioral interventions. *Clin Diabetes Endocrinol* 2021;**7**:1–9. doi:10.1186/s40842-021-00120-z
10  Horsch A, Kang JS, Vial Y, *et al.* Stress exposure and psychological stress responses are related to glucose concentrations during pregnancy. *Br J Health Psychol* 2016;**21**:712–29. doi:10.1111/bjhp.12197
11  Twedt R, Bradley M, Deiseroth D, *et al.* Sleep Duration and Blood Glucose Control in Women With Gestational Diabetes Mellitus. *Obstet Gynecol* 2015;**126**:326–31. doi:10.1126/science.1249098.Sleep
12  Craig L, Sims R, Glasziou P, *et al.* Women's experiences of a diagnosis of gestational diabetes mellitus: A systematic review. *BMC Pregnancy Childbirth* 2020;**20**:1–15. doi:10.1186/s12884-020-2745-1

13  Parsons J, Ismail K, Amiel S, *et al.* Perceptions among women with gestational diabetes. *Qual Health Res* 2014;**24**:575–85. doi:10.1177/1049732314524636
14  Persson M, Winkvist A, Mogren I. 'From stun to gradual balance'- women's experiences of living with gestational diabetes mellitus. *Scand J Caring Sci* 2010;**24**:454–62. doi:10.1111/j.1471-6712.2009.00735.x
15  Carolan M, Gill GK, Steele C. Women's experiences of factors that facilitate or inhibit gestational diabetes self-management. *BMC Pregnancy Childbirth* 2012;**12**. doi:10.1186/1471-2393-12-99
16  Arabin B, Baschat AA. Pregnancy: An Underutilized Window of Opportunity to Improve Long-term Maternal and Infant Health - An Appeal for Continuous Family Care and Interdisciplinary Communication. *Front Pediatr* 2017;**5**. doi:10.3389/fped.2017.00069
17  Xie W, Dai P, Qin Y, *et al.* Effectiveness of telemedicine for pregnant women with gestational diabetes mellitus: an updated meta-analysis of 32 randomized controlled trials with trial sequential analysis. *BMC Pregnancy Childbirth* 2020;**20**:198. doi:10.1186/s12884-020-02892-1
18  Miremberg H, Ben-ari T, Betzer T, *et al.* The impact of a daily smartphone-based feedback system among women with gestational diabetes on compliance, glycemic control, satisfaction, and pregnancy outcome: a randomized controlled trial. *The American Journal of Obstetrics & Gynecology* 2018;**218**:453.e1-453.e7. doi:10.1016/j.ajog.2018.01.044
19  Yang P, Lo W, He Z lin, *et al.* Medical nutrition treatment of women with gestational diabetes mellitus by a telemedicine system based on smartphones. *Journal of Obstetrics and Gynaecology Research* 2018;**44**:1228–34. doi:10.1111/jog.13669
20  Borgen I, Småstuen MC, Jacobsen AF, *et al.* Effect of the Pregnant+ smartphone application in women with gestational diabetes mellitus : a randomised controlled trial in Norway. *BMJ Open* 2019;**9**:1–12. doi:10.1136/bmjopen-2019-030884
21  Mackillop L, Hirst JE, Bartlett KJ, *et al.* Comparing the Efficacy of a Mobile Phone-Based Blood Glucose Management System With Standard Clinic Care in Women With Gestational Diabetes: Randomized Controlled Trial. *J Med Internet Res* 2018;**20**:e71. doi:10.2196/mhealth.9512
22  Kytö M, Strömberg L, Tuomonen H, *et al.* Behavior Change Apps for Gestational Diabetes Management: Exploring Desirable Features. *Int J Hum Comput Interact* 2022;**38**:1095–112. doi:10.1080/10447318.2021.1987678
23  Kytö M, Koivusalo S, Ruonala A, *et al.* Behavior Change App for Self-management of Gestational Diabetes: Design and Evaluation of Desirable Features. *JMIR Hum Factors* 2022;**9**:e36987. doi:10.2196/36987
24  Mamykina L, Heitkemper EM, Smaldone AM, *et al.* Personal discovery in diabetes self-management: Discovering cause and effect using self-monitoring data. *J Biomed Inform* 2017;**76**:1–8. doi:10.1016/j.jbi.2017.09.013
25  Ayobi A, Marshall P, Cox AL, *et al.* Quantifying the body and caring for the mind: Self-tracking in multiple sclerosis. In: *Conference on Human Factors in Computing Systems - Proceedings*. 2017. 6889–901. doi:10.1145/3025453.3025869
26  Li I, Dey AK, Forlizzi J. Understanding My Data, Myself: Supporting Self-Reflection with Ubicomp Technologies. In: *Ubicomp '11 Proceedings of the 13th ACM international conference on Ubiquitous computing*. 2011. 405–14. doi:10.1146/annurev.py.07.090169.000411
27  MacLeod H, Tang A, Carpendale S. Personal informatics in chronic illness management. *Proceedings - Graphics Interface* 2013;:149–56.

28   Mishra SR, Klasnja P, Woodburn JMD, *et al.* Supporting coping with Parkinson's disease through self-tracking. In: *CHI*. 2019. 1–16. doi:10.1145/3290605.3300337
29   Schroeder J, Chung CF, Epstein DA, *et al.* Examining self-tracking by people with migraine: Goals, needs, and opportunities in a chronic health condition. In: *DIS 2018 - Proceedings of the 2018 Designing Interactive Systems Conference*. 2018. 135–48. doi:10.1145/3196709.3196738
30   Årsand E, N. T, G. O, *et al.* Mobile phone-based self-management tools for type 2 diabetes: the few touch application. *J Diabetes Sci Technol* 2010;**4**:328–36.
31   Årsand E, Muzny M, Bradway M, *et al.* Performance of the first combined smartwatch and smartphone diabetes diary application study. *J Diabetes Sci Technol* 2015;**9**:556–63. doi:10.1177/1932296814567708
32   Weatherall J, Paprocki Y, Meyer TM, *et al.* Sleep tracking and exercise in patients with type 2 diabetes mellitus (Step-D): Pilot study to determine correlations between fitbit data and patient-reported outcomes. *J Med Internet Res* 2018;**20**. doi:10.2196/mhealth.8122
33   Maugeri A, Barchitta M, Agodi A. How Wearable Sensors Can Support the Research on Foetal and Pregnancy Outcomes: A Scoping Review. *J Pers Med* 2023;**13**. doi:10.3390/jpm13020218
34   Raj S, Lee J, Garrity A, *et al.* Clinical Data in Context: Towards Sensemaking Tools for Interpreting Personal Health Data. *Proc ACM Interact Mob Wearable Ubiquitous Technol* 2019;**3**:1–20. doi:10.1145/3314409
35   Karkar R, Zia J, Vilardaga R, *et al.* A framework for self-experimentation in personalized health. *Journal of the American Medical Informatics Association* 2016;**23**:440–8. doi:10.1093/jamia/ocv150
36   Mamykina L, Mynatt ED, Davidson PR, *et al.* MAHI: Investigation of social scaffolding for reflective thinking in diabetes management. In: *CHI'08*. 2008. 477–86.
37   Raj S, Toporski K, Garrity A, *et al.* 'My blood sugar is higher on the weekends': Finding a Role for Context and Context-Awareness in the Design of Health Self-Management Technology. In: *CHI*. 2019. 1–13.
38   Li I, Dey A, Forlizzi J. A stage-based model of personal informatics systems. In: *Conference on Human Factors in Computing Systems - Proceedings*. 2010. 557–66. doi:10.1145/1753326.1753409
39   Draffin CR, Alderdice FA, Mccance DR, *et al.* Exploring the needs, concerns and knowledge of women diagnosed with gestational diabetes : A qualitative study. *Midwifery* 2016;**40**:141–7. doi:10.1016/j.midw.2016.06.019
40   Carolan M. Women' s experiences of gestational diabetes self-management: A qualitative study. *Midwifery* 2013;**29**:637–45. doi:10.1016/j.midw.2012.05.013
41   Hood M, Wilson R, Corsica J, *et al.* What do we know about mobile applications for diabetes self-management? A review of reviews. *J Behav Med* 2016;**39**:981–94. doi:10.1007/s10865-016-9765-3
42   Skar JB, Garnweidner-Holme LM, Lukasse M, *et al.* Women's experiences with using a smartphone app (the Pregnant+ app) to manage gestational diabetes mellitus in a randomised controlled trial. *Midwifery* 2018;**58**:102–8. doi:10.1016/j.midw.2017.12.021
43   Surendran S, Lim CS, Koh GCH, *et al.* Women's usage behavior and perceived usefulness with using a mobile health application for gestational diabetes mellitus: Mixed-methods study. *Int J Environ Res Public Health* 2021;**18**. doi:10.3390/ijerph18126670
44   Jovanovic L. The Role of Continuous Glucose Monitoring in Gestational Diabetes Mellitus. *Diabetes Technol Ther* 2000;**2**:11–5.

45  Lane A, Mlynarczyk M, de Veciana M, *et al.* Real-time continuous glucose monitoring in gestational diabetic pregnancies: a randomized controlled trial. *Am J Obstet Gynecol* 2019;**220**:S68–9. doi:10.1016/j.ajog.2018.11.094

46  Scott EM, Bilous RW, Kautzky-Willer A. Accuracy, User Acceptability, and Safety Evaluation for the FreeStyle Libre Flash Glucose Monitoring System When Used by Pregnant Women with Diabetes. *Diabetes Technol Ther* 2018;**20**:180–8. doi:10.1089/dia.2017.0386

47  Voormolen DN, DeVries JH, Sanson RME, *et al.* Continuous glucose monitoring during diabetic pregnancy (GlucoMOMS): A multicentre randomized controlled trial. *Diabetes Obes Metab* 2018;**20**:1894–902. doi:10.1111/dom.13310

48  Wei Q, Sun Z, Yang Y, *et al.* Effect of a CGMS and SMBG on Maternal and Neonatal Outcomes in Gestational Diabetes Mellitus: A Randomized Controlled Trial. *Sci Rep* 2016;**6**:1–9. doi:10.1038/srep19920

49  Rooksby J, Rost M, Morrison A, *et al.* Personal Tracking as Lived Informatics. In: *CHI*. 2014. 1163–72. doi:10.1145/2556288.2557039

50  Catalano PM, Huston L, Amini SB, *et al.* Longitudinal changes in glucose metabolism during pregnancy in obese women with normal glucose tolerance and gestational diabetes mellitus. *Am J Obstet Gynecol* 1999;**180**:903–16.

51  Hermsen S, Frost J, Renes RJ, *et al.* Using feedback through digital technology to disrupt and change habitual behavior: A critical review of current literature. *Comput Human Behav* 2016;**57**:61–74. doi:10.1016/j.chb.2015.12.023

52  Kooiman TJM, de Groot M, Hoogenberg K, *et al.* Self-tracking of Physical Activity in People with Type 2 Diabetes: A Randomized Controlled Trial. *CIN - Computers Informatics Nursing* 2018;**36**:340–9. doi:10.1097/CIN.0000000000000443

53  Chan KL, Chen M. Effects of social media and mobile health apps on pregnancy care: Meta-analysis. *JMIR Mhealth Uhealth* 2019;**7**:e11836: 1-13. doi:10.2196/11836

54  Peyton T, Poole E, Reddy M, *et al.* 'Every pregnancy is different': Designing mHealth interventions for the pregnancy ecology. In: *Proceedings of the ACM Conference on Designing Interactive Systems (DIS 2014)*. 2014. 577–86. doi:bpnh

55  Rigla M, Martínez-Sarriegui I, García-Sáez G, *et al.* Gestational Diabetes Management Using Smart Mobile Telemedicine. *J Diabetes Sci Technol* 2018;**12**:260–4. doi:10.1177/1932296817704442

56  Böhm A-K, Jensen ML, Sørensen MR, *et al.* Real-World Evidence of User Engagement With Mobile Health for Diabetes Management : Longitudinal Observational Study. *JMIR Mhealth Uhealth* 2020;**8**:e22212. doi:10.2196/22212

57  Evenson KR, Wen F. National trends in self-reported physical activity and sedentary behaviors among pregnant women: NHANES 1999-2006. *Prev Med (Baltim)* 2010;**50**:123–8. doi:10.1016/j.ypmed.2009.12.015

58  Boudreaux BD, Hebert EP, Hollander DB, *et al.* Validity of Wearable Activity Monitors during Cycling and Resistance Exercise. *Med Sci Sports Exerc* 2018;**50**:624–33. doi:10.1249/MSS.0000000000001471

59  Penders J, Altini M, van Hoof C, *et al.* Wearable Sensors for Healthier Pregnancies. *Proceedings of the IEEE* 2015;**103**:179–91. doi:10.1109/JPROC.2014.2387017

60  Runkle J, Sugg M, Boase D, *et al.* Use of wearable sensors for pregnancy health and environmental monitoring: Descriptive findings from the perspective of patients and providers. *Digit Health* 2019;**5**:205520761982822. doi:10.1177/2055207619828220

61	Bailey TS, Ahmann A, Brazg R, *et al.* Accuracy and acceptability of the 6-day enlite continuous subcutaneous glucose sensor. *Diabetes Technol Ther* 2014;**16**:277–83. doi:10.1089/dia.2013.0222
62	Vähä-Ypyä H, Vasankari T, Husu P, *et al.* Validation of cut-points for evaluating the intensity of physical activity with accelerometry-based Mean Amplitude Deviation (MAD). *PLoS One* 2015;**10**:1–13. doi:10.1371/journal.pone.0134813
63	Vähä-Ypyä H, Husu P, Suni J, *et al.* Reliable recognition of lying, sitting, and standing with a hip-worn accelerometer. *Scand J Med Sci Sports* 2018;**28**:1092–102. doi:10.1111/sms.13017
64	Husu P, Suni J, Vähä-Ypyä H, *et al.* Objectively measured sedentary behavior and physical activity in a sample of Finnish adults: A cross-sectional study. *BMC Public Health* 2016;**16**:1–11. doi:10.1186/s12889-016-3591-y
65	Husu P, Tokola K, Vähä-Ypyä H, *et al.* Physical Activity, Sedentary Behavior, and Time in Bed Among Finnish Adults Measured 24/7 by Triaxial Accelerometry. *J Meas Phys Behav* 2021;**4**:163–73. doi:10.1123/jmpb.2020-0056
66	Fokkema T, Kooiman TJM, Krijnen WP, *et al.* Reliability and validity of ten consumer activity trackers depend on walking speed. *Med Sci Sports Exerc* 2017;**49**:793–800. doi:10.1249/MSS.0000000000001146
67	Byrne NM, Groves AM, McIntyre HD, *et al.* Changes in resting and walking energy expenditure and walking speed during pregnancy in obese women. *American Journal of Clinical Nutrition* 2011;**94**:819–30. doi:10.3945/ajcn.110.009399
68	Parak J, Korhonen I. Accuracy of Firstbeat Bodyguard HR monitor. 2015. https://assets.firstbeat.com/firstbeat/uploads/2015/11/white_paper_bodyguard2_final.pdf
69	Norton K, Norton L, Sadgrove D. Position statement on physical activity and exercise intensity terminology. *J Sci Med Sport* 2010;**13**:496–502. doi:10.1016/j.jsams.2009.09.008
70	Shin J, Lee S, Gong T, *et al.* MyDJ: Sensing Food Intakes with an Attachable on Your Eyeglass Frame. In: *Conference on Human Factors in Computing Systems - Proceedings*. Association for Computing Machinery 2022. doi:10.1145/3491102.3502041
71	Hassannejad H, Matrella G, Ciampolini P, *et al.* Automatic diet monitoring: a review of computer vision and wearable sensor-based methods. *Int J Food Sci Nutr* 2017;**68**:656–70. doi:10.1080/09637486.2017.1283683
72	Yang YJ, Kim MK, Hwang SH, *et al.* Relative validities of 3-day food records and the food frequency questionnaire. *Nutr Res Pract* 2010;**4**:142. doi:10.4162/nrp.2010.4.2.142
73	Venkatesh V, Morris MG, Davis GB, *et al.* User Acceptance of Information Technology: Toward a Unified View. *MIS Quarterly* 2003;**27**:425–78. doi:10.2307/30036540
74	Gonder-frederick LA, Shepard JA, Grabman JH, *et al.* Psychology, Technology, and Diabetes Management. *American Psychologist* 2016;**71**:577–89.
75	Gale NK, Heath G, Cameron E, *et al.* Using the framework method for the analysis of qualitative data in multi-disciplinary health research. *BMC Med Res Methodol* 2013;**13**:1. doi:10.1186/1471-2288-13-117
76	Corbin J, Strauss A. *Basics of qualitative research: Techniques and procedures for developing grounded theory*. SAGE Publications 2015.
77	Finnish institute of health and welfare. Medical Birth register. https://thl.fi/en/web/thlfi-en/statistics-and-data/data-and-services/register-descriptions/newborns. 2019.
78	Yu W, Wu N, Li L, *et al.* A Review of Research Progress on Glycemic Variability and Gestational Diabetes. *Diabetes Metab Syndr Obes* 2020;**Volume 13**:2729–41. doi:10.2147/dmso.s261486

79  Mottola MF, Davenport MH, Ruchat SM, *et al.* Canadian Guideline for Physical Activity throughout Pregnancy. *Journal of Obstetrics and Gynaecology Canada* 2018;**40**:1528–37. doi:10.1016/j.jogc.2018.07.001
80  Pien GW, Schwab RJ. Sleep Disorders During Pregnancy. *Sleep* 2004;**27**:1405–17.
81  Speranza G, Verlato G, Albiero A. Autonomic Changes During Pregnancy Assessment by Spectral Heart Rate Variability Analysis. *J Electrocardiol* 1998;**31**:101–9.
82  Kim HG, Cheon EJ, Bai DS, *et al.* Stress and heart rate variability: A meta-analysis and review of the literature. *Psychiatry Investig* 2018;**15**:235–45. doi:10.30773/pi.2017.08.17
83  Conover WJ. *Practical nonparametric statistics*. 2nd ed. New York: : Wiley 1980.
84  Carolan-Olah M, Steele C, Krenzin G. Development and initial testing of a GDM information website for multi-ethnic women with GDM. *BMC Pregnancy Childbirth* 2015;**15**:1–9. doi:10.1186/s12884-015-0578-0
85  Chung C-F, Wang Q, Schroeder J, *et al.* Identifying and Planning for Individualized Change: Patient-Provider Collaboration Using Lightweight Food Diaries in Healthy Eating and Irritable Bowel Syndrome. *Proc ACM Interact Mob Wearable Ubiquitous Technol* 2019;**3**:1–27. doi:10.1145/3314394
86  Hayase M, Shimada M, Seki H. Sleep quality and stress in women with pregnancy-induced hypertension and gestational diabetes mellitus. *Women and Birth* 2014;**27**:190–5. doi:10.1016/j.wombi.2014.04.002
87  Avery MelissaD, Walker AJ. Acute effect of exercise on blood glucose and insulin levels in women with gestational diabetes. *Journal of Maternal-Fetal Medicine* 2001;**10**:52–8. doi:10.1080/714904296
88  Chen H, Fang X, Wong TH, *et al.* Physical Activity during Pregnancy: Comparisons between Objective Measures and Self-Reports in Relation to Blood Glucose Levels. *Int J Environ Res Public Health* 2022;**19**. doi:10.3390/ijerph19138064
89  Choe EK, Lee NB, Lee B, *et al.* Understanding quantified-selfers' practices in collecting and exploring personal data. In: *Conference on Human Factors in Computing Systems - Proceedings*. 2014. 1143–52. doi:10.1145/2556288.2557372
90  Rapp A, Cena F. Personal informatics for everyday life: How users without prior self-tracking experience engage with personal data. *International Journal of Human Computer Studies* 2016;**94**:1–17. doi:10.1016/j.ijhcs.2016.05.006
91  Schroeder J, Karkar R, Murinova N, *et al.* Examining Opportunities for Goal-Directed Self-Tracking to Support Chronic Condition Management. *Proc ACM Interact Mob Wearable Ubiquitous Technol* 2019;**3**. doi:10.1145/3369809.Examining
92  Kraaij V, Garnefski N. Cognitive, Behavioral and Goal Adjustment Coping and Depressive Symptoms in Young People with Diabetes: A Search for Intervention Targets for Coping Skills Training. *J Clin Psychol Med Settings* 2015;**22**:45–53. doi:10.1007/s10880-015-9417-8
93  Harrison AL, Taylor NF, Frawley HC, *et al.* Women with gestational diabetes mellitus want clear and practical messages from credible sources about physical activity during pregnancy: a qualitative study. *J Physiother* 2019;**65**:37–42. doi:10.1016/j.jphys.2018.11.007
94  Desai PM, Mitchell EG, Hwang ML, *et al.* Personal Health Oracle: Explorations of Personalized Predictions in Diabetes Self-Management. In: *CHI19*. 2019. 1–13. doi:10.1145/3290605.3300600
95  Dempsey PC, Owen N, Biddle SJH, *et al.* Managing sedentary behavior to reduce the risk of diabetes and cardiovascular disease. *Curr Diab Rep* 2014;**14**. doi:10.1007/s11892-014-0522-0
96  Matthews CE, Hagströmer M, Pober D, *et al.* Best Practices for Physical Activity Monitors in Population-Based Research. *Med Sci Sports Exerc* 2012;**44**:1–17. doi:10.1249/MSS.0b013e3182399e5b.BEST

97   Williams L, Hayes GR, Guo Y, *et al.* HCI and mHealth Wearable Tech: A Multidisciplinary Research Challenge. In: *Extended Abstracts of the 2020 CHI Conference on Human Factors in Computing Systems - CHI '20*. 2020. 1–7. doi:10.1145/3334480.3375223

98   Grym K, Niela-Vilén H, Ekholm E, *et al.* Feasibility of smart wristbands for continuous monitoring during pregnancy and one month after birth. *BMC Pregnancy Childbirth* 2019;**19**:1–9. doi:10.1186/s12884-019-2187-9

99   Andersen TO, Langstrup H, Lomborg S. Experiences With Wearable Activity Data During Self-Care by Chronic Heart Patients: Qualitative Study. *J Med Internet Res* 2020;**22**:e15873. doi:10.2196/15873

100  Mills GH. Sleeping positions adopted by pregnant women of more than 30 weeks gestation. *Anaesthesia* 1994;**49**:249–50.

101  Pais S, Parry D, Rush E, *et al.* Data integration for mobile wellness apps to support treatment of GDM. In: *Proceedings of the Australasian Computer Science Week Multiconference on - ACSW '16*. 2016. 1–7. doi:10.1145/2843043.2843382

102  Lin T, Gal A, Mayzel Y, *et al.* Non-Invasive Glucose Monitoring: A Review of Challenges and Recent Advances. *Curr Trends Biomed Eng Biosci* 2017;**6**:1–8. doi:10.19080/ctbeb.2017.06.555696

103  Greenwood DA, Gee PM, Fatkin KJ, *et al.* A Systematic Review of Reviews Evaluating Technology-Enabled Diabetes Self-Management Education and Support. *J Diabetes Sci Technol* 2017;**11**:1015–27. doi:10.1177/1932296817713506

104  Muuraiskangas S, Mattila E, Kyttälä P, *et al.* User Experiences of a Mobile Mental Well-Being Intervention Among Pregnant Women. In: Serino S, Matic A, Giakoumis D, *et al.*, eds. *Communications in Computer and Information Science*. 2016. 140–9. doi:10.1007/978-3-319-32270-4

105  West P, van Kleek M, Giordano R, *et al.* Common Barriers to the Use of Patient-Generated Data Across Clinical Settings. In: *CHI*. 2018. 1–13. doi:10.1145/3173574.3174058

106  Soleimani H, Subbaswamy A, Saria S. Treatment-response models for counterfactual reasoning with continuous-time, continuous-valued interventions. In: *Uncertainty in Artificial Intelligence - Proceedings of the 33rd Conference, UAI 2017*. 2017.

107  Peleg M, Shahar Y, Quaglini S, *et al.* MobiGuide: a personalized and patient-centric decision-support system and its evaluation in the atrial fibrillation and gestational diabetes domains. *User Model User-adapt Interact* 2017;**27**:159–213. doi:10.1007/s11257-017-9190-5

108  Hamzeh M, Oliver KL. 'Because I am muslim, I cannot wear a swimsuit': Muslim girls negotiate participation opportunities for physical activity. *Res Q Exerc Sport* 2012;**83**:330–9. doi:10.1080/02701367.2012.10599864

109  Sempionatto JR, Montiel VRV, Vargas E, *et al.* Wearable and Mobile Sensors for Personalized Nutrition. ACS Sens. 2021;**6**:1745–60. doi:10.1021/acssensors.1c00553

110  Kytö M, Markussen LT, Marttinen P, *et al.* Comprehensive self-tracking of blood glucose and lifestyle with a mobile application in the management of gestational diabetes: a study protocol for a randomised controlled trial (eMOM GDM study). *BMJ Open* 2022;**0**. doi:10.1136/bmjopen-2022-066292